\documentclass[aps,prd,preprintnumbers,superscriptaddress,nofootinbib,floatfix,twocolumn,notitlepage]{revtex4-2}
\usepackage{graphicx}  
\usepackage{dcolumn}   
\usepackage{bm}        
\usepackage{epsfig,amsmath,amssymb,verbatim,mathrsfs,array,layout,textcomp,amssymb,latexsym,slashed}
\usepackage{xcolor}
\usepackage{siunitx}
\usepackage[colorlinks=true,citecolor=blue,urlcolor=blue,linktocpage=true,
linkcolor=blue]{hyperref}
\usepackage[utf8]{inputenc}
\usepackage{multirow}
\usepackage{amsthm}
\usepackage{thmtools}
\usepackage[capitalize,noabbrev]{cleveref}

\usepackage[normalem]{ulem}

\def\beq{\begin{equation}}
\def\eeq{\end{equation}}
\def\beqa{\begin{eqnarray}}
\def\eeqa{\end{eqnarray}}

\begin{document}

\title{ 
 Natural Phantom Dark Energy from a $\mathbb{Z}_N$--Axion}
\preprint{LAPTH-041/26}

%
\author{C\'edric Delaunay}
\email{cedric.delaunay@lapth.cnrs.fr}
\affiliation{Laboratoire d'Annecy de Physique Th\'eorique, CNRS--USMB, 9 chemin de Bellevue, 74940 Annecy, France}
\author{Admir Greljo}
\email{admir.greljo@unibas.ch}
\affiliation{Department of Physics, University of Basel, Klingelbergstrasse 82, CH-4056 Basel, Switzerland}
%


\begin{abstract}
We present a technically natural microscopic realization of apparent phantom dark energy based on an axion coupled to $N$ copies of two-flavor dark QCD related by a $\mathbb{Z}_N$ exchange symmetry. The symmetry exponentially suppresses the axion vacuum potential, naturally generating the dark-energy scale, while reheating into a single dark sector induces a controlled breaking of $\mathbb{Z}_N$ that simultaneously restores the physical axion periodicity, sets the dark-matter abundance, and drives the late-time dark-energy dynamics. The selected sector contains dark-pion dark matter, whose finite density initially traps the axion away from its vacuum minimum. As the Universe expands and the dark-pion density redshifts away, the axion is released and rolls on the reheating-induced vacuum potential, generating an effective phantom crossing without tuned cancellations. We identify a viable parameter region that simultaneously reproduces the observed dark-matter relic abundance and dark-energy scale, satisfies cosmological and astrophysical constraints, and qualitatively reproduces the DESI preference for an evolving dark-energy equation of state.
\end{abstract}

\maketitle

%

\section{Introduction}
\label{sec:introduction}

``\textit{All you need is $\Lambda$, $\Lambda$ is all you need.}''
For more than two decades~\cite{SupernovaSearchTeam:1998fmf}, cosmological observations appeared to support this picture. A cosmological constant (CC), combined with cold dark matter (CDM), provides an excellent description of the
expansion history of the Universe. Within the resulting \(\Lambda\)CDM
framework, dark energy (DE) is interpreted as a constant vacuum energy density with equation of state \(w_\Lambda=-1\), in remarkable agreement with data from supernovae (SNe), the cosmic microwave background (CMB), and large-scale structure (LSS).

Recent baryon acoustic oscillation (BAO) measurements from the Dark Energy
Spectroscopic Instrument (DESI), however, may hint at a departure from this simple picture~\cite{DESI:2024mwx, DESI:2025zgx}. When analyzed jointly with existing SNe~\cite{Brout:2022vxf, Rubin:2023jdq, DES:2024jxu} and
CMB~\cite{Planck:2018nkj, Planck:2019nip, Efstathiou:2019mdh,
Carron:2022eyg, Rosenberg:2022sdy, ACT:2023kun, ACT:2023oei} datasets, and
using the two-parameter Chevallier--Polarski--Linder (CPL) parametrization
\(w_{\rm DE}=w_0+w_a(1-a)\), the DESI data exhibit a \(\sim3\sigma\)
preference for evolving dark energy, with the preferred region lying in the
quadrant \(w_0>-1\), \(w_a<0\), and \(w_0+w_a<-1\). This corresponds to an
apparent crossing of the phantom divide in the recent past, although the
significance and detailed redshift dependence remain dataset- and
parametrization-dependent, see e.g.~\cite{DESI:2025fii, Wolf:2025jlc, Gialamas:2025pwv, Toomey:2025xyo, Lee:2025pzo}.

A time-varying DE component is not, by itself, particularly surprising from a
theoretical perspective. Such behavior can arise from the late-time evolution
of a slowly rolling scalar field~\cite{Ratra:1987rm, Wetterich:1987fm},
commonly referred to as quintessence~\cite{Caldwell:1997ii}. The tentative
preference for phantom behavior~\cite{Caldwell:1999ew}, \(w_{\rm DE}<-1\), is considerably more
challenging to accommodate. Within general relativity (GR), an isolated
canonical scalar field, or more generally an isolated fluid obeying the
null-energy condition, satisfies \(\rho+p\geq 0\), and therefore
\(w\equiv p/\rho\geq -1\)~\cite{Qiu:2007fd, Ludwick:2017tox, Moghtaderi:2025cns, Caldwell:2025inn}. Consequently, a confirmed detection of
\(w_{\rm DE}<-1\) would point to physics beyond canonical, minimally coupled
quintessence. One interesting possibility is that interactions between DE and
other cosmic fluids lead to an effective phantom equation of
state~\cite{Das:2005yj, Smith:2024ibv, Khoury:2025txd, Bedroya:2025fwh, Wang:2025znm, Guedezounme:2025wav, Chen:2025ywv, LaPenna:2026avs, Antusch:2026ldp}.
In this case, the fundamental degrees of freedom need not violate the
null-energy condition; rather, the phantom behavior is inferred by an observer
who analyzes the expansion history in terms of noninteracting DM and
DE.

Among the many proposals for dynamical DE~\cite{Tsujikawa:2013fta},
axion-like fields~\cite{Frieman:1995pm, Kim:1998kx, Kim:1999dc, Liu:2025bss} are
particularly compelling because they provide a technically natural explanation
for an ultralight scalar degree of freedom. In this framework, the shift
symmetry of the pseudo Nambu--Goldstone boson (pNGB) \(\varphi\), arising from
a spontaneously broken global symmetry, protects its ultralight mass from
radiative corrections. Similarly to the QCD axion, a nontrivial potential
\(V(\varphi)\) can emerge from explicit breaking of the shift symmetry due to a
mixed anomaly between the global symmetry and a confining gauge group,
\begin{equation}
    V(\varphi)
    =
    \Lambda^4
    \left[
        1-\cos\left(\frac{\varphi}{f}\right)
    \right] ,
\end{equation}
where \(\Lambda\) is related to the confinement scale and \(f\) is the decay
constant of \(\varphi\). For an initial displacement of order \(f\), the field
remains frozen by Hubble friction until late times, and then begins to roll.
The observed DE density requires \(\Lambda\sim{\cal O}({\rm meV})\). For the
field to roll at present, its mass must satisfy \(m_\varphi\sim H_0\), with
\(H_0\sim{\cal O}(10^{-33}\,{\rm eV})\) the present-day Hubble parameter.
Consequently, \(f\sim \Lambda^2/m_\varphi\sim{\cal O}(10^{18}\,{\rm GeV})\).
The resulting decay constant lies close to the Planck scale, reflecting the
well-known coincidence between the onset of cosmic acceleration and the
present Hubble scale.

A particularly attractive realization of an effective phantom crossing arises when the axion couples directly to CDM through a field-dependent DM mass \(m_{\rm DM}(\varphi)\). In this case, the DM energy density no longer
redshifts exactly as \(\rho_{\rm DM}\propto a^{-3}\), where \(a\) is the scale
factor of the Universe, leading to an effective phantom-like DE
equation of state~\cite{Das:2005yj}. Such couplings arise naturally when DM
itself is a bound state of the same confining gauge group responsible for the
DE potential. The DM background then contributes a finite-density correction
to the axion potential,
\begin{equation}
\label{eq:DEpot}
    V(\varphi)
    =
    \Lambda^4
    \left[
        1+
        \left(
            \frac{\sigma n_{\rm DM}}{\Lambda^4}
            -1
        \right)
        \cos\left(\frac{\varphi}{f}\right)
    \right]
    +\cdots ,
\end{equation}
where \(\sigma>0\) denotes the change of the bound-state mass between the
finite-density and vacuum configurations, and the ellipsis represents
corrections nonlinear in \(n_{\rm DM}\), depending on the DM kinetic energy and
self-interactions. This is analogous to the case of the QCD axion in nuclear
matter, where \(\sigma\) is the pion--nucleon sigma term
\(\sigma_{\pi N}\equiv m_q(\partial m_N/\partial m_q)\), controlling the
dependence of the nucleon mass \(m_N\) on the quark masses
\cite{Cohen:1991nk, Hook:2017psm}.

At first sight, the structure of Eq.~\eqref{eq:DEpot} seems to allow for the
desired cosmological evolution. At early times, when
\(n_{\rm DM}\gg \Lambda^4/\sigma\), the finite-density term dominates and
drives the field toward \(\varphi=\pi f\), independently of its initial
condition after inflation. As the Universe expands and the DM density dilutes,
the vacuum contribution eventually takes over and shifts the minimum back to
\(\varphi=0\). The subsequent late-time motion of the axion field realizes
dynamical DE and can lead to apparent phantom crossing~\cite{Khoury:2025txd}.

This simple picture, however, faces two closely related obstructions. The
first is a control problem. The change of minimum must occur while the dark
matter is still a dilute gas, so that the expansion of the finite-density
energy in powers of \(n_{\rm DM}\) remains reliable. In a generic confining
theory, however, the vacuum axion potential and the DM mass are
controlled by the same strong scale. The critical density at which the
finite-density contribution competes with the vacuum potential is then
parametrically comparable to the density at which the dilute expansion breaks
down. The sign flip of the axion potential therefore occurs precisely where
the linear term in Eq.~\eqref{eq:DEpot} can no longer be trusted. This is the
familiar situation for the QCD axion in a background of nucleons, where the
relevant density is set by the saturation scale associated with the
pion--nucleon sigma term~\cite{Cohen:1991nk, Hook:2017psm}.

The second obstruction is cosmological. If the same confinement scale sets
both the DE potential and the DM mass, then the observed
DE scale would suggest \(m_{\rm DM}\sim{\cal O}({\rm meV})\). Such a
light relic is relativistic around matter--radiation equality and is excluded
as the dominant DM. One may try to separate the two scales by
introducing light dark quarks, so that the axion potential is chirally
suppressed while the confinement scale, and hence the DM mass, remains
larger. For two flavors this suppression is controlled by
\(z_{ud}=m_u/m_d\). Requiring \(m_{\rm DM}\gtrsim 10\,{\rm keV}\) to evade
warm-DM (WDM) bounds then demands an extremely small quark-mass ratio,
\[
    z_{ud} \lesssim 10^{-28}
    \left(\frac{m_{\rm DM}}{10\,{\rm keV}}\right)^4 .
\]
Although technically natural in the chiral sense, this would amount to a
severe flavor puzzle in the dark sector.

A more economical way to obtain an exponentially small axion potential is to
use a discrete \(\mathbb{Z}_N\) exchange symmetry among \(N\) identical
confining sectors~\cite{Hook:2018jle, DiLuzio:2021pxd}. The axion couples to the \(N\) copies
with phases shifted by \(2\pi k/N\). In the \(\mathbb{Z}_N\)-symmetric limit,
the leading harmonics cancel in the sum over copies, and the remaining vacuum
potential is suppressed as
\[
    V_{\mathbb{Z}_N}
    \sim
    z_{ud}^N \,\Lambda^4
    \cos\left(\frac{N\varphi}{f}\right) .
\]
The small DE scale can then be obtained with an order-one hierarchy among
dark-quark masses, at the price of a moderately large number of copies. For
example, \(z_{u d}\approx0.5 \, (0.1)\) requires \(N \gtrsim 100 \,(30)\). This also lowers the
density at which the finite-density term overcomes the vacuum potential,
allowing the minimum flip to occur well inside the dilute regime.

However, this solution introduces a new obstruction. The same
\(\mathbb{Z}_N\) symmetry that suppresses the vacuum energy also reduces the
periodicity of the vacuum potential from \(2\pi f\) to \(2\pi f/N\)
\cite{Hook:2018jle}. The axion then moves only over a field range of order
\(f/N\) near the would-be finite-density minimum. For large \(N\), the
late-time dynamics are therefore strongly suppressed, and the model becomes
observationally indistinguishable from a CC. In addition,
the curvature of the \(\cos(N\varphi/f)\) potential is enhanced by \(N^2\),
potentially pushing the required decay constant above the Planck scale.

The central observation of this work is that this last obstruction is removed
if the DM population itself selects one of the \(N\) copies. A
finite-density contribution shared equally among all copies would respect the
\(\mathbb{Z}_N\) symmetry and would inherit the reduced \(2\pi f/N\)
periodicity. By contrast, if only one copy contains the relic dark matter, the
finite-density energy explicitly breaks \(\mathbb{Z}_N\) and depends on the
axion with the original \(2\pi f\) periodicity. The axion can then be held near
the density-induced minimum at early times and later roll over an
\({\cal O}(f)\) field range once the density redshifts away.

We show that this copy selection need not be imposed by hand. It can arise
from reheating. A reheaton coupled to the Standard Model and to a single dark
copy populates only that copy after inflation. The same coupling softly breaks
the \(\mathbb{Z}_N\) exchange symmetry in the vacuum, shifting the dark-quark
masses in the selected sector and generating the leading \(2\pi f\)-periodic
axion vacuum potential. Thus the two ingredients required for the cosmological
mechanism---the DM abundance and the vacuum potential that releases
the axion at late times---originate from the same spurion.

This logic also clarifies why the identity of the DM matters. A
baryonic realization, along the lines of earlier finite-density axion models,
is not minimal in the present setup. In a symmetric thermal history, dark
baryons annihilate efficiently into the lighter dark mesons and are typically
subdominant. An asymmetric baryonic abundance can be arranged, but reproducing
the observed DM density while maintaining an appreciable
DM--axion coupling requires a large primordial dark asymmetry,
parametrically much larger than the baryon asymmetry of the Standard Model.
Moreover, the accompanying dark-meson population must either be sufficiently
depleted or made unstable without conflicting with BBN, CMB, and
dark-radiation constraints. We therefore focus on the more economical
possibility in which the relic DM is itself composed of the lightest
dark mesons. In the two-flavor theory these are the dark pions, whose finite
density provides the axion-dependent contribution responsible for the
early-time trapping of the field.

The resulting construction has three essential ingredients. First, a
\(\mathbb{Z}_N\)-symmetric axion sector exponentially suppresses the vacuum
energy while allowing the dark confinement scale, and hence the dark-pion
mass, to remain well above the meV scale. Second, reheating into a
single copy provides a controlled and cosmologically motivated source of
\(\mathbb{Z}_N\) breaking. Third, the selected copy contains a relic population
of dark pions whose axion-dependent mass drives an effective energy exchange
between DM and DE at late times. We derive the axion potential including
both the vacuum and finite-density contributions, compute the resulting
background evolution, and identify the region in which the dark-pion relic
abundance, the dark-energy scale, BBN, warm-dark-matter bounds,
self-interactions, and chiral control can be simultaneously satisfied.

The paper is organized as follows. In \cref{sec:ZNaxion}, we introduce the
\(\mathbb{Z}_N\)-axion construction and its UV completion. In
\cref{sec:DEpotential}, we derive the axion potential in the presence of
the selected dark-pion density. In \cref{sec:Phantom}, we study the
resulting background evolution and the effective phantom crossing. In
\cref{sec:growth}, we discuss the impact on the growth of matter perturbations. In \cref{sec:ZNbreaking}, we show how reheating into a single
copy generates the required \(\mathbb{Z}_N\) breaking, and in \cref{sec:DMrelic}, we compute the dark-pion relic abundance. The viable
parameter space is summarized in \cref{sec:alltogether}, and we conclude in \cref{sec:concl}.

\section{A $\mathbb{Z}_N$--Axion Model}\label{sec:ZNaxion}


Consider $N$ identical copies of a dark sector comprised of a confining gauge group $G_D\equiv{\rm SU}(N_c)$ and two vector-like fermions $q = u, d$ in the fundamental representation, which all couple to a single axion field. The Lagrangian of this effective field theory (EFT) is 
\begin{align}\label{eq:Leff}
\mathcal{L}_{\rm dark}^{\rm eff}\supset&\sum_{k=0}^{N-1} \left[\left(\frac{\varphi}{f} +\frac{2\pi k}{N}\right)\frac{\alpha_D}{4\pi}{\rm tr}\,G_{k} \widetilde G_k-\sum_{q=u,d}m_q\bar q_kq_k\right]\,,
\end{align}
where $G_k$ are the gluon field strengths; the canonical kinetic terms are understood. This theory exhibits a discrete $\mathbb{Z}_N$ symmetry acting as 
\begin{align}
    q_k,G_k\to q_{k+1},G_{k+1}\,,\quad \varphi\to \varphi +\frac{2\pi f}{N}\,,
\end{align}
with $q_{N-1}\to q_0$, $G_{N-1}\to G_0$, and $\varphi/f$ is $2\pi$ periodic. The dark quark masses $m_u$ and $m_d$ are taken to be much smaller than the confinement scale, giving rise to pion triplets as the lightest dark states. A universal dark $\bar\theta$ angle can be absorbed into a constant shift of $\varphi$.\\

The Lagrangian in~\cref{eq:Leff} can be obtained, for instance, from a
KSVZ completion~\cite{Kim:1979if, Shifman:1979if}. In the minimal
realization, one introduces in each copy a vector-like fermion $Q_k$ in
the fundamental representation of ${\rm SU}(N_c)_k$, together with a
single Peccei--Quinn (PQ) scalar 
\begin{equation}
\Phi=\frac{f_\Phi+\rho}{\sqrt2}e^{i\varphi/f_\Phi}.
\end{equation}
The PQ charges obey
\begin{equation}
{\rm PQ}(\Phi)= {\rm PQ}(Q_{L k})-{\rm PQ}(Q_{R k})=1,
\end{equation}
so that bare Dirac masses for $Q_k$ are forbidden, while the Yukawa
interactions
\begin{equation}
\mathcal{L}_{\rm KSVZ}\supset
-\sum_{k=0}^{N-1}
y_Q e^{2\pi i k/N}\Phi\,\bar Q_{Lk}Q_{Rk}
+{\rm h.c.}
\end{equation}
are allowed. These interactions are invariant under the cyclic $\mathbb Z_N$ symmetry
\begin{equation}
Q_k\to Q_{k+1},\qquad
\Phi\to e^{2\pi i/N}\Phi.    
\end{equation}
After PQ breaking, and choosing a basis in which $y_Q$ is real, the heavy fermion masses are
\begin{equation}
m_{Qk}=
\frac{y_Q f_\Phi}{\sqrt2}
e^{i(\varphi/f_\Phi+2\pi k/N)}. 
\end{equation}
Removing these phases by chiral rotations of $Q_k$ generates the
couplings in~\cref{eq:Leff}. For one fundamental KSVZ fermion per copy
with ${\rm PQ}(Q_L)-{\rm PQ}(Q_R)=1$, the mixed
PQ-${\rm SU}(N_c)_k^2$ anomaly coefficient is unity for each dark gauge
group. Thus each sector separately would produce an axion potential with
domain-wall number $N_{\rm DW}=1$, and the normalization of
\cref{eq:Leff} corresponds to $f=f_\Phi$. The full theory contains $N$ such contributions, whose sum produces a
$\mathbb Z_N$-invariant contribution to the axion potential with period
$2\pi f/N$, as discussed below.\\

In KSVZ models~\cite{Kim:1979if, Shifman:1979if, Palavric:2026vej}, the light dark quarks appearing in~\cref{eq:Leff} are
uncharged under PQ symmetry, ${\rm PQ}(q_k)=0$. Their masses may arise
from a separate structure unconstrained by PQ. For example, in each copy
one may introduce a dark gauged SU(2) under which
$q_{Lk}=(u_{Lk},d_{Lk})^T$ transforms as a doublet, while
$u_{Rk}$ and $d_{Rk}$ are singlets.\footnote{Alternatively, one may gauge a dark chiral U(1). In analogy with the SM hypercharge, anomaly cancellation may then require introducing additional fermions.} The $\mathbb{Z}_N$ symmetry acts as
$S_k\to S_{k+1}$ on the corresponding scalar doublets. Once
\(S_k\) acquires a vacuum expectation value, this SU(2) is fully
Higgsed and the Yukawa interactions
\begin{equation}
\mathcal{L}\supset  \sum_{k=0}^{N-1} \left[ - y_u\bar q_{Lk}\widetilde S_k u_{Rk}
-y_d\bar q_{Lk}S_k d_{Rk} \right]
+{\rm h.c.}\,, 
\end{equation}
generate the dark-quark masses $m_u$ and $m_d$. The Witten anomaly~\cite{Witten:1982fp}
requires the total number of left-handed $SU(2)$ doublets to be even.
Since each generation comes with $N_c$ dark-color copies, the condition
is $n_gN_c \in 2\mathbb Z$, thus for odd $N_c$, an even number $n_g$ of
such generations is required, with only the lightest retained in the EFT.

As discussed in \cref{sec:ZNbreaking}, the reheaton field $\phi$ softly breaks the $\mathbb Z_N$ symmetry through the operator $\kappa_D \phi S_0^\dagger S_0$. This induces a small shift in $\langle S_0\rangle$ relative to the other copies and therefore a corresponding shift in the quark masses of the $k=0$ dark-QCD sector. For the discussion of the axion potential, we parametrize this explicit breaking as
\begin{equation}\label{eq:ebdef}
m_{u_0,d_0}=m_{u,d}\left(1+\epsilon_b\right),
\end{equation}
where $\epsilon_b \ll 1$ denotes the breaking spurion. The origin of $\epsilon_b$ will be discussed in \cref{sec:ZNbreaking}.

\section{Dark energy potential}
\label{sec:DEpotential}

We now derive the axion potential relevant to late-time evolution.  The starting point is the leading-order chiral potential in each dark-QCD copy.  For a fixed value of the axion field, it is useful to keep the neutral pion direction $\alpha_k$ explicit.  We write
\begin{equation}
    U_k = \exp\left(i \alpha_k \tau^3\right),
    \qquad
    \theta_k \equiv \frac{\varphi}{f}+\frac{2\pi k}{N},
\end{equation}
so that the two-flavor chiral potential in the $k$-th copy is
\begin{equation}
    V_{\chi,k}(\theta_k,\alpha_k)
    =
    - B_0 f_\pi^2
    \left[
        m_d \cos \alpha_k
        +
        m_u \cos(\alpha_k-\theta_k)
    \right] .
\end{equation}
Minimizing over $\alpha_k$ gives \begin{equation} 
V_{\chi,k}^{\rm min}(\theta_k) = -m_\pi^2 f_\pi^2\, r(\theta_k), 
\end{equation} where 
\begin{equation} 
m_\pi^2 \equiv B_0(m_u+m_d), \qquad z_{ud} \equiv \frac{m_u}{m_d}, 
\end{equation} 
and 
\begin{equation} 
r(\theta) \equiv \sqrt{ 1- \frac{4 z_{ud}}{(1+z_{ud})^2} \sin^2\frac{\theta}{2} }\,. 
\end{equation}
In the $\mathbb Z_N$-symmetric limit, the total vacuum potential is obtained by summing over the $N$ copies,
\begin{equation}
    V_{\mathbb Z_N}(\Theta)
    =
    - m_\pi^2 f_\pi^2
    \sum_{k=0}^{N-1}
    r\left(\Theta+\frac{2\pi k}{N}\right),
    \quad
    \Theta \equiv \frac{\varphi}{f}.
\end{equation}
The leading $\Theta$-dependent term in this sum appears first at order $z_{ud}^N$,
\begin{equation}
    V_{\mathbb Z_N}(\Theta)
    \simeq
     m_\pi^2 f_\pi^2
    \frac{(-z_{ud})^N}{\sqrt{\pi N}} \cos(N\Theta)\,.
\end{equation}
This is the exponentially protected vacuum contribution to the axion potential.

As discussed in \cref{sec:ZNbreaking}, the reheaton field $\phi$ softly breaks the $\mathbb Z_N$ symmetry. Parametrizing this explicit breaking as in \cref{eq:ebdef}.  the corresponding contribution to the axion potential is
\begin{equation}
    V_b(\Theta)
    =
    \epsilon_b m_\pi^2 f_\pi^2
    \left[
        1-r(\Theta)
    \right].
\end{equation}
For the soft breaking to dominate over the residual $\mathbb Z_N$-symmetric vacuum term in the curvature around the origin, we require parametrically
\begin{equation}\label{eq:ZNconditionN}
    \epsilon_b
    \gtrsim
    \frac{N^{3/2}}{\sqrt{\pi}}\,
    z_{ud}^{N-1}(1+z_{ud})^2 .
\end{equation}

The selected $k=0$ copy also contains the relic population of dark pions.  Since these pions are non-relativistic and dilute, their leading contribution to the energy density is
\begin{equation}
    \rho_\pi(\Theta)=n_\pi\, m_\pi(\Theta),
\end{equation}
where $n_\pi$ is the number density and $m_\pi(\Theta)$ is the pion mass in the axion background. Strictly, this term should be added before integrating out the neutral pion direction.  In the background $(\Theta,\alpha)$,
\begin{equation}
    m_\pi^2(\Theta,\alpha)
    =
    B_0
    \left[
        m_d\cos\alpha
        +
        m_u\cos(\alpha-\Theta)
    \right],
\end{equation}
and therefore
\begin{equation}
    V_d(\Theta,\alpha)
    =
    n_\pi
         m_\pi(\Theta,\alpha)\,.
\end{equation}
We define the pion-restoring potential
\begin{equation}
    \Delta V_\pi(\Theta,\alpha)
    \equiv
    V_{\chi,0}(\Theta,\alpha)
    -
    V_{\chi,0}^{\rm min}(\Theta),
\end{equation}
which vanishes at the vacuum pion minimum for fixed \(\Theta\).  Up to an additive constant, the axion potential in the presence of the pion density is then
\begin{equation}
    V_{\rm DE}(\Theta)
    =
    \min_\alpha
    \left[
        \Delta V_\pi(\Theta,\alpha)
        +
        V_d(\Theta,\alpha)
    \right]+ V_{b}(\Theta)+ V_{\mathbb Z_N}(\Theta).
\label{eq:VDE-exact-alpha}
\end{equation}

In the regime of interest,
\begin{equation}
    n_\pi \sim \epsilon_b m_\pi f_\pi^2,
\end{equation}
the density-induced shift of the pion minimum is only $\mathcal O(\epsilon_b)$, so its effect on the minimized potential starts at $\mathcal O(\epsilon_b^2 m_\pi^2 f_\pi^2)$.  Thus, to leading order, the breaking and density terms can be evaluated at the vacuum pion minimum. After neglecting $V_{\mathbb Z_N}(\Theta)$ assuming \cref{eq:ZNconditionN} is satisfied, the full, density-dependent axion is
\begin{equation}
    V_{\rm DE}(\Theta)
    \simeq
    V_0+\Lambda_b 
    \Big[
        1-r(\Theta)
    \Big]
    +
    n_\pi m_\pi        \sqrt{r(\Theta)},
    \label{eq:VDE-leading}
\end{equation}
where $\Lambda_b\equiv \epsilon_b m_\pi^2 f_\pi^2$. Here, $V_0$ denotes an additive constant contribution to the vacuum energy. Its value is not predicted within our framework and is assumed to be fixed by whatever mechanism ultimately resolves the infamous CC problem, for example~\cite{Weinberg:1988cp}. 

The explicit-breaking term selects the vacuum minimum at $\Theta=0$, while the finite-density contribution favors configurations with smaller dark-pion masses and therefore stabilizes the field near $\Theta=\pi$ when the pion density is sufficiently large. The subsequent evolution is controlled by the competition between these two terms.

\section{Phantom divide crossing}\label{sec:Phantom}

Identifying the $\mathbb{Z}_N$  axion $\varphi$ as the field responsible for late-time DE, we now study its cosmological evolution and identify the parameter space leading to a crossing of the phantom divide similar to DESI reconstruction. 
Because the dark-pion mass depends on $\varphi$, this evolution induces an exchange of energy between DE and DM, causing the energy density of the latter to deviate from the standard  scaling as $a^{-3}$. 
An observer interpreting the resulting cosmological evolution within a framework of noninteracting DE would therefore infer an effective equation of state~\cite{Das:2005yj,Khoury:2025txd},
\begin{align}\label{eq:wDE}
w_{\rm DE}^{\rm eff}=w_\varphi\left[1+\frac{\rho_{\rm DM}}{\rho_\varphi}\left[A(\varphi)-1\right]\right]^{-1}
\end{align}
where $w_\varphi\equiv p_\varphi/\rho_\varphi$, with $p_\varphi\equiv \dot{\varphi}^{ 2}/2-V$ and $\rho_\varphi\equiv \dot{\varphi}^{ 2}/2+V$ denoting the pressure and energy density of $\varphi$, respectively. Here $\rho_{\rm DM}\equiv m_\pi(\varphi_0) n_\pi\propto a^{-3}$, with $\varphi_0$ the present value of the axion field, is a reference DM energy density with the standard redshift scaling of pressureless matter, and  
\begin{equation}
    A(\varphi)\equiv \frac{m_\pi(\varphi)}{m_\pi(\varphi_0)}\,.
\end{equation}
Therefore, phantom crossing ($w_{\rm DE}^{\rm eff}<-1$) is obtained for $A(\varphi)<1$, that is, when the dark-pion mass increases at late times.\\

The origin of this effective equation of state can be understood from the coupled evolution of the DM and DE components. As a result of the energy transfer with $\varphi$, the dark-pion energy density obeys the modified continuity equation, 
\begin{equation}\label{eq:continuityeq}
\rho_\pi'+3\mathcal{H}\rho_\pi- \varphi'\beta(\varphi)\rho_\pi=0\,,
\end{equation}
where primes denote derivatives with respect to conformal time, $\mathcal{H}\equiv a'/a$ is the conformal Hubble parameter, and $\beta(\varphi)\equiv d\log m_\pi/d\varphi$.  
The evolution of the axion field is governed by the Klein-Gordon equation,
\begin{equation}\label{eq:KGphi}
\varphi''+2\mathcal{H}\varphi'+a^2\left[\frac{dV}{d\varphi}+\beta(\varphi)\rho_\pi\right]=0\,,
\end{equation}
where the second term inside the brackets arises from the 
$\varphi$-dependence of the dark-pion mass and represents the backreaction of the dark-pion density on the axion dynamics.\\

The axion field may be initially displaced from the minimum of its potential on cosmological scales if the PQ symmetry is broken before inflation. Inflation then stretches a single Hubble patch, characterized by a random initial value of the field $\varphi_i$, to encompass the entire observable Universe, resulting in an effectively homogeneous axion field. To determine its subsequent evolution, we solve the coupled system in~\cref{eq:continuityeq} and \cref{eq:KGphi} starting deep in the radiation era at $z_i=10^7$, with initial conditions  $\varphi(z_i)=\varphi_i\sim \mathcal{O}(f)$ and  $\varphi'(z_i)=0$ for the axion field. 

The initial radiation density is fixed by the observed CMB photon temperature~\cite{Fixsen_2009}, assuming three relativistic neutrino species and neglecting the small contribution of neutrino masses to the late-time energy budget. Furthermore, we adopt the $\Lambda$CDM values~\cite{Planck:2018vyg} for the baryon abundance and Hubble at $z_i$, while the initial dark-pion abundance is fixed through the flatness condition. As a result, the expansion history at early times remains essentially identical to that of the standard $\Lambda$CDM cosmology, with deviations appearing only when the axion field begins to evolve dynamically at late times.\\

Four parameters controls the dynamics of the axion field: the  vacuum energy offset $V_0$, the overall scale of the vacuum potential $\Lambda_b$, the quark mass ratio $z_{ud}$, and the axion decay constant $f$. At early times, the finite-density contribution dominates the axion potential and generates an effective axion mass 
\begin{equation}
m_a^2=\frac{z_{ud}  m_\pi n_\pi}{2f^2}\left[1+\mathcal{O}(z_{ud})\right]\,.
\end{equation}
Consequently, for $f\lesssim \sqrt{3z_{ud}}M_{\rm P}/2$, the axion field starts evolving from its inflationary value toward the density-induced attractor at $\varphi=\pi f$ before matter-radiation equality, rendering the subsequent evolution qualitatively insensitive to the precise initial value $\varphi_i$. For the representation benchmark point considered below, the field typically begins rolling during the radiation era and settles near the attractor shortly after recombination. 

As the Universe expands and the dark-pion density redshifts as $n_\pi\propto a^{-3}$, the finite-density contribution progressively weakens. Once  $n_\pi\leqslant2\Lambda_b/m_\pi$, the vacuum potential becomes dominant and the axion mass asymptotes to the constant value,
\begin{equation}
m_a^2\simeq\frac{z_{ud}\Lambda_b}{f^2}\left[1+\mathcal{O}(z_{ud})\right]\,. 
\end{equation}
At this transition,
the energy density stored in the axion field relative to that of the dark pions is $\rho_\Theta/\rho_\pi\simeq 1+V_0/(2\Lambda_b)$, implying that the Universe is already DE dominated for positive $V_0$. The disappearance of the finite-density potential then releases the axion field from the attractor, allowing it to roll toward the true vacuum at $\varphi=0$, provided $f\lesssim \sqrt{3z_{ud}/(2+V_0/\Lambda_b)} M_{\rm P}$. The redshift at which this transition occurs is determined by $\Lambda_b$, and is approximately given by 
\begin{equation}
z\simeq -1+1.35\left(\frac{\Lambda_b}{H_0^2M_{\rm P}^2}\right)^{1/3}\,,
\end{equation}
where $\sqrt{H_0M_{\rm P}}\simeq 1.86\,{\rm meV}$. Since the axion field is released from a position close to the maximum of the vacuum potential at $\varphi=\pi f$, its subsequent evolution, and hence the onset of the effective phantom crossing, is typically delayed. The duration of this delay is controlled by the small residual velocity at the time of release, which can be adjusted by the initial displacement $\varphi_i$ generated during inflation.\\

\begin{figure}
    \centering
\includegraphics[width=1\linewidth]{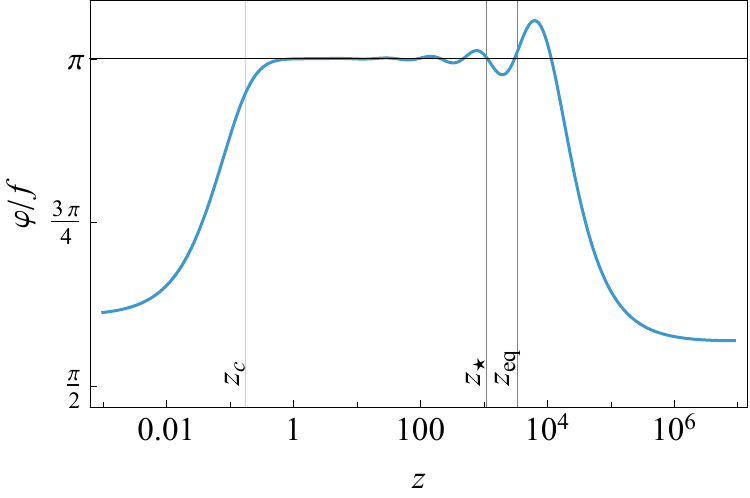}
    \caption{Evolution of the axion field $\varphi$ as a function of redshift for a benchmark set of parameters, obtained by numerically solving the coupled DM--DE evolution equations. The field is initially attracted to the finite-density minimum at $\varphi=\pi f$, then slowly rolls toward the vacuum minimum at $\varphi=0$ once the finite-density contribution becomes subdominant. The vertical lines at $z_{\rm eq}\approx3320$, $z_\star\approx1090$, and $z_c\approx0.17$ denote the epochs of matter--radiation equality, recombination, and the effective phantom crossing predicted by the benchmark solution, respectively.}
    \label{fig:axionsolution}
\end{figure}

In~\cref{fig:axionsolution}, we present the numerical solution for the axion field evolution for a representative benchmark choice of the model parameters:
\begin{eqnarray}
    V_0=\ 0.37H_0^2M_{\rm P}^2\,,\quad& \Lambda_b=19H_0^2M_{\rm P}^2\,,&\quad z_{ud}=0.05\,,\nonumber\\
    f=\ 0.065M_{\rm P}\,,\quad& \varphi_i=0.57\pi f\,.&
\end{eqnarray}
The benchmark point is chosen such that the angular acoustic scale $\theta_\star$ agrees with the Planck measurement~\cite{Planck:2018vyg} within one standard deviation, while the total matter abundance remains within $2\%$ of its $\Lambda$CDM value.

\begin{figure}[t]
    \centering
\includegraphics[width=1\linewidth]{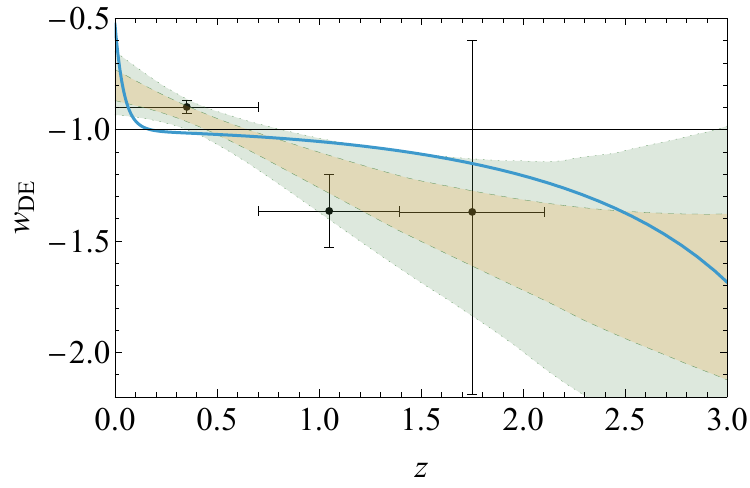}
    \caption{Effective dark energy equation of state as a function of redshift, obtained for a representative benchmark choice of the model parameters by numerically solving the coupled DM-DE evolution equations (blue line). Black points with error bars correspond to the binned DESI DR2 reconstruction~\cite{DESI:2024mwx}, while the shaded dark and light bands indicate the $68\%$ and $95\%$ confidence regions, respectively, obtained from Gaussian process reconstructions of $w_{\rm DE}(z)$ using the combined DESI, CMB, and Union3 datasets~\cite{DESI:2025fii}.}
    \label{fig:wDEplot}
\end{figure}

The corresponding effective DE equation of state, defined in~\cref{eq:wDE}, is shown in~\Cref{fig:wDEplot}. As the field rolls toward the vacuum minimum at late time, the dark-pion mass continuously increases, causing the DM energy density to redshift more slowly than the standard $a^{-3}$ scaling and giving rise to an effective phantom crossing, qualitatively reproducing the trend suggested by DESI reconstruction. The benchmark solution exhibits a relatively late phantom crossing at $z_c\simeq0.17$. This reflects that the axion is released from a field value close to the maximum of the vacuum potential at $\varphi=\pi f$, where the potential is particularly flat, thereby delaying its subsequent evolution. We present additional background diagnostics for the benchmark solution in Appendix~\ref{app:DEdiagnostics}.\\   

\section{Matter perturbations}\label{sec:growth}

The coupling between the axion and dark pions also modifies the evolution of cosmological perturbations. Since the dark-pion mass depends on the local value of the axion field, spatial fluctuations of $\varphi$ induce fluctuations in the DM mass, giving rise to an effective fifth-force interaction within the dark sector~\cite{Archidiacono:2022iuu, Bottaro:2024pcb}. Conversely, perturbations in the dark-pion density modify the finite-density contribution to the axion potential, thereby sourcing axion fluctuations. As a result, the linear perturbations of the DM and DE sectors become dynamically coupled, leading to departures from the standard $\Lambda$CDM growth history. 

Working in the conformal Newtonian gauge and asusming negligible anisotropic stress, the evolution of the dark-pion density contrast $\delta_\pi(\boldsymbol{r},t)\equiv\rho_\pi(\boldsymbol{r},t)/\bar \rho_\pi(t)-1$, velocity divergence $\theta_\pi$, and axion perturbation $\delta\varphi(\boldsymbol{x},t)\equiv\varphi(\boldsymbol{x},t)-\bar\varphi(t)$ in Fourier space with comoving wavenumber $\boldsymbol{k}$ is governed at linear order by
\begin{align}
    \delta_\pi'+\theta_\pi-3\Psi'-\delta\varphi'\beta(\bar\varphi)-\bar\varphi'\delta\varphi\frac{d\beta}{d\bar\varphi}&=0\,,\\
    \theta_\pi'+\left[\mathcal{H}+\beta(\bar\varphi)\bar\varphi'\right]\theta_\pi-k^2\left[\Psi+\beta(\bar\varphi)\delta\varphi\right]&=0\,,
\end{align}
where barred quantities denote homogeneous background values, $\Psi$ is the Newtonian gravitational potential, and $\delta\varphi$ satisfies the perturbed Klein-Gordon equation
\begin{widetext}
\begin{equation}
\delta\varphi''+2\mathcal{H}\delta\varphi'-4\bar\varphi'\Psi'+\delta\varphi\left[k^2+a^2\left(\frac{d^2V}{d\bar\varphi^2}+\bar\rho_\pi\frac{d\beta}{d\bar\varphi}\right)\right]
+2\Psi a^2\left[\frac{dV}{d\bar\varphi}+\bar\rho_\pi\beta(\bar\varphi)\right]+a^2\bar\rho_\pi \delta_\pi \beta(\bar\varphi)=0\,.
\end{equation}
\end{widetext}
The term proportional to \(\beta\delta\varphi\) is the scalar fifth force,
while the term \(\beta\bar\varphi'\theta_\pi\) is an additional drag induced by the time dependence of the dark-pion mass. In the
phantom-crossing regime of interest, the dark-pion mass typically increases at
late times, so \(\beta\bar\varphi'>0\), and this term acts as an extra
friction on the dark-pion velocity.

These equations are supplemented by the usual perturbation equations for
photons and baryons, together with the Einstein equations for the
metric perturbation \(\Psi\). The expressions above assume that the dark pions
are already nonrelativistic and that their pressure and velocity dispersion
can be neglected on the large scales relevant for the late-time growth
analysis.

To solve the linear perturbation equations, we impose adiabatic initial conditions at $z_i=10^7$ coming from inflation. At this redshift, all Fourier modes relevant for the computation of the matter power spectrum satisfy $k\ll\mathcal H$ and are therefore outside the cosmological horizon. The primordial scalar perturbations from inflation are assumed to be nearly scale invariant, with amplitude $A_s$ and spectral index $n_s$ fixed to their Planck 2018 best-fit values~\cite{Planck:2018vyg}. Since $\bar\varphi'(z_i)=0$, adiabaticity further requires that the axion perturbations vanish initially, yielding $\delta\varphi(z_i)=\delta\varphi'(z_i)=0$.
The remaining perturbation variables are therefore initialized according to the standard adiabatic conditions of the $\Lambda$CDM cosmology. Since the axion field is frozen at early times, the evolution of cosmological perturbations remains essentially indistinguishable from that of $\Lambda$CDM until the onset of the late-time axion dynamics.\\

The impact of the DM--DE interaction on structure formation can be quantified through $f\sigma_8(z)$, where $f\equiv d\ln D/d\ln a$ is the linear growth rate, with $D(a)$ the linear growth factor of matter perturbations, and $\sigma_8(z)$ is the root-mean-square amplitude of matter fluctuations smoothed over spheres of radius $8\,h^{-1}\,\mathrm{Mpc}$. The quantity $f\sigma_8$, which is directly constrainted by redshift-space distortions (RSD)~\cite{Beutler:2012px,Howlett:2014opa,Blake:2013nif,WiggleZ:2012sek,Pezzotta:2016gbo,Okumura:2015lvp} and peculiar velocity surveys (PVS)~\cite{Said:2020epb,Boruah:2019icj,Carrick:2015xza,Huterer:2016uyq,Turner:2022mla}, provides a particularly robust probe of departures from the standard $\Lambda$CDM growth history. 

\begin{figure}
    \centering
\includegraphics[width=1\linewidth]{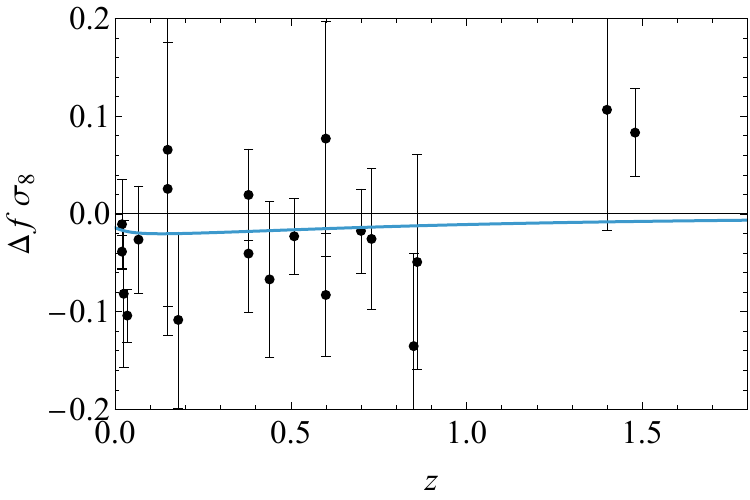}
    \caption{Evolution of the linear growth observable $\Delta f\sigma_8(z)\equiv f\sigma_8(z)-f\sigma_8^{\Lambda{\rm CDM}}(z)$  for a representative benchmark choice of the axion DE parameters (blue line), obtained by solving the coupled background and linear perturbation equations,  relative to the prediction of the $\Lambda$CDM model~\cite{Euclid:2025bxg}. The black points with error bars correspond to current redshift-space distortion measurements~\cite{Beutler:2012px,Howlett:2014opa,Blake:2013nif,WiggleZ:2012sek,Pezzotta:2016gbo,Okumura:2015lvp} and peculiar velocity surveys~\cite{Said:2020epb,Boruah:2019icj,Carrick:2015xza,Huterer:2016uyq,Turner:2022mla}. The DE--DM interaction induced by the axion-dependent dark-pion mass typically suppresses the growth of matter perturbations relative to $\Lambda$CDM by a few percent over the redshift range probed by current observations, while remaining compatible with existing data.}
    \label{fig:growth}
\end{figure}

In~\cref{fig:growth}, we present the shift 
\begin{equation}
    \Delta f\sigma_8(z)\equiv f\sigma_8(z)-f\sigma_8^{\Lambda{\rm CDM}}(z)
\end{equation}
in the growth observable $f\sigma_8(z)$ for the representative benchmark choice of the axion DE parameters discussed in~\cref{sec:Phantom}, relative to the $\Lambda$CDM prediction~\cite{Euclid:2025bxg}. We find that the DM--DE interaction typically suppresses the growth of matter perturbations by a few percent relative to $\Lambda$CDM over the redshift range probed by current observations.
This suppression is not caused by the scalar fifth force, which tends to enhance DM clustering on scales below the Compton wavelength of the axion field~\cite{Archidiacono:2022iuu}. Instead, it results from the combined effect of the modified Hubble flow due to the time
dependence of the dark-pion mass and the additional velocity drag
\(\beta\bar\varphi'\theta_\pi\) during the epoch in which the effective
phantom behavior is generated. In particular, matter domination occurs slightly later than in $\Lambda$CDM, thereby reducing the time available for the growth of matter perturbations~\cite{Khoury:2025txd}.

\section{$\mathbb{Z}_N$  breaking from reheating}
\label{sec:ZNbreaking}

We propose a minimal realization in which a single source of $\mathbb Z_N$ symmetry breaking simultaneously generates the additional contribution to the axion potential required for phantom dark energy and determines the dark matter relic abundance.

Specifically, we assume that only one copy of the dark sector, which we label $k=0$, is reheated after inflation together with the SM bath. This is achieved by introducing a real scalar reheaton field $\phi$ with interactions
\begin{equation}
\mathcal L_\phi \supset
-\frac{m_\phi^2}{2}\phi^2
-\phi\left(
\kappa_{\rm SM}H^\dagger H
+\kappa_D S_0^\dagger S_0
\right),
\end{equation}
where $H$ denotes the SM Higgs doublet and $S_0$ the dark Higgs field of the $k=0$ sector. The absence of analogous couplings to the dark Higgs fields of the remaining copies explicitly breaks the $\mathbb Z_N$ symmetry. Since the breaking enters through the dimension-three operator $\phi S_0^\dagger S_0$, it is soft and therefore suppressed in the UV. The trilinear couplings are also small spurions of higher-spin symmetry of the $\phi$ field~\cite{Hook:2023pba}, and one expects $\kappa_{D,{\rm SM}} \ll m_\phi$.\\

At the end of reheating, the ratio of energy densities stored in the dark and visible sectors is determined by the reheaton branching fractions,
$\rho_D/\rho_{\rm SM}=\Gamma_{\phi \to D}/\Gamma_{\phi\to {\rm SM}}$. For a sufficiently large $m_\phi$, the reheaton decays promptly, with the partial decay widths given approximately by
\begin{equation}\label{eq:decreh}
    \Gamma_{\phi\to {\rm SM},D}=\frac{\kappa^2_{{\rm SM},D}}{8\pi  m_\phi}\,,
\end{equation} 
to decay into the SM and the $k=0$ dark sector, respectively. We restrict our discussion to the parameter space in which $1 \to 2$ decay is kinematically accessible. Following reheating, each sector rapidly thermalizes internally while remaining thermally decoupled from the others. The resulting temperature ratio  $\xi\equiv T'/T$, where $T$ and $T'$ denote the visible- and dark-sector temperatures, is therefore fixed by the ratio of energy densities deposited into the two sectors, 
\begin{align}\label{eq:xirh}
    \xi_{\rm rh}=\left[\frac{g_\star^{\rm SM}(T_{\rm rh}) \rho_D}{g_\star^D(T_{\rm rh}')\rho_{\rm SM}}\right]^{1/4}\,,
\end{align}
where $g_\star^{{\rm SM},D}$ denote the numbers of relativistic degrees of freedom in the SM plasma and the $k=0$ dark sector. When $T'_{\rm rh}>T_c'$ where $T_c'\sim f_\pi$ is the temperature of the dark confinement phase transition, the dark sector is reheated in the deconfined phase and the number of relativistic degrees of freedom is
$g_\star^D(T'_{\rm rh})=2(N_c^2-1)+7N_c=37$
for $N_c=3$ and assuming two light quark flavors. Conversely, if the dark sector is reheated below the confinement scale but above the dark-pion mass, $m_\pi\lesssim T'_{\rm rh}<T_c'$,
the only relativistic degrees of freedom are the dark pions and
$g_\star^D(T'_{\rm rh})=g_\pi=3$.
 This yields
\begin{equation}\label{eq:Thratio}
    \xi_{\rm rh}=
    \alpha_{\rm rh}
    \sqrt{\frac{\kappa_D}{\kappa_{\rm SM}}}
    \left[
    \frac{g_\star^{\rm SM}(T_{\rm rh})}{10.75}
    \right]^{1/4},
\end{equation}
where $\alpha_{\rm rh}\simeq 0.73\,(1.4)$ when reheating is above (below) the dark confinement scale, assuming $T_{\rm rh}\sim1-10\,\mathrm{MeV}$.\\

As discussed in \cref{sec:DMrelic}, reproducing the observed DM relic abundance typically requires $\kappa_D < \kappa_{\rm SM}$, such that the decay width into the SM dominates the Hubble rate at the end of reheating. The SM reheating temperature is then defined by $\Gamma_{\phi\to {\rm SM}}=H(T_{\rm rh})$, which gives
\begin{equation}\label{eq:Trh}
    T_{\rm rh}=4.6\,{\rm MeV}\sqrt{\frac{m_\phi}{M_{\rm P}}}\left(\frac{\kappa_{\rm SM}/m_\phi}{10^{-20}}\right)\left[\frac{10.75}{g_\star^{\rm SM}(T_{\rm rh})}\right]^{1/4},
\end{equation}
where $M_{\rm P}\approx 2.4\times 10^{18}\,$GeV is the reduced Planck mass. Requiring standard Big Bang nucleosynthesis (BBN) imposes $T_{\rm rh}\gtrsim 5\,$MeV, which translates into a lower bound on the reheaton coupling to the SM,  $\kappa_{\rm SM}/m_\phi\gtrsim 10^{-20}\sqrt{M_{\rm P}/m_\phi}$.\\

The same $\mathbb{Z}_N$--breaking coupling $\kappa_D$ also backreacts on the axion potential. It does so by shifting the vacuum expectation value (VEV) of $S_0$, and hence the quark mass in the $k=0$ dark sector, relative to those in the other copies. The relevant scalar potential is
\begin{align}
V(\phi,S_0,H)=&(-\mu_S^2+\kappa_D\phi)S_0^\dagger S_0+\lambda_S(S^\dagger_0 S_0)^2\nonumber\\
&+(-\mu^2+\kappa_{\rm SM}\phi)H^\dagger H+\lambda(H^\dagger H)^2\nonumber\\
&+\frac{m_\phi^2}{2}\phi^2\pm \kappa_{\rm SM}\frac{\Lambda_{\rm UV}^2}{16\pi^2}\phi\,,
\end{align}
where $H$ is the SM Higgs doublet and $\mu^2_{S},\,\mu^2$ are positive. The last term parametrizes the one-loop reheaton tadpole induced by Higgs loops ($\kappa_{\rm SM} > \kappa_D$)~\cite{Banerjee:2025zcd}. Its precise magnitude and sign depend on the ultraviolet completion at the cutoff scale $\Lambda_{\rm UV}$.   Minimizing this potential, this tadpole induces a small expectation value for $\phi$, which in turn shifts the VEV of $S_0$ relative to those of $S_{k\neq0}$. This VEV shift induces the parameter $\epsilon_b$ introduced in~\cref{eq:ebdef},
$\epsilon_b\simeq(\langle S_0\rangle -\langle S_{k\neq 0}\rangle)/\langle S_{k\neq 0}\rangle$, one finds
\begin{equation}
    \epsilon_b\simeq
    \frac{\kappa_D\kappa_{\rm SM}}{2 m_\phi^2}\left[\pm\frac{\Lambda_{\rm UV}^2}{16\pi^2\mu_S^2}+\frac{v^2}{\mu_S^2}\right]\,,
\end{equation}
where $v\equiv \sqrt{\mu^2/(2\lambda)}\ll \Lambda_{\rm UV}$ is the SM Higgs VEV. In view of \cref{eq:VDE-leading}, the relevant question is what the smallest achievable $\epsilon_b$ is in this model. Neglecting possible tuned cancelations against other contributions in the tadpole, EFT consistency implies $\mu_S \lesssim \Lambda_{\rm UV}/(4\pi)$. We also restrict ourselves to $m_\phi \gtrsim 2 \mu_{ S}$, so that $\phi \to S_0^\dagger S_0$ is kinematically allowed and \cref{eq:decreh} applies. As a result,
\begin{equation}
|\epsilon_b| \gtrsim \frac{\kappa_D\kappa_{\rm SM}}{2 m_\phi^2}\,.    
\end{equation}

In this estimate, we have neglected the dependence of $\Lambda_D$ on $\langle S\rangle$ due to threshold corrections to the running of the dark gauge coupling. This is a small effect in ordinary QCD, where approximately $\Lambda_{\rm QCD}\propto v^{0.25}
$~\cite{Agrawal:1997gf, Agrawal:1998xa}, and the effect should be smaller in the dark sector under consideration, which contains only two quark flavors above the confinement scale. In the following, we assume $\epsilon_b>0$, so that the vacuum axion potential is minimized at $\Theta=0$. Finally, the Higgs portal contribution $H^\dagger HS^\dagger S$, generated by tree-level $\phi$ exchange, is $\propto \kappa_D\kappa_{\rm SM}/m_\phi^2$ and is far too small to thermalize the two sectors. The absence of other contributions to this term can be justified by separate Poincaré symmetry for decoupled sectors.\\ 

Crucially, the parameter $\epsilon_b$ is not an independent deformation of the theory, but is instead directly tied to the cosmic history. The coupling $\kappa_{\rm SM}$ fixes the initial SM radiation energy density through the reheating temperature in \cref{eq:Trh}, while the ratio $\kappa_D/\kappa_{\rm SM}$ fixes the initial temperature of the dark sector in \cref{eq:Thratio}. As shown in~\cref{sec:DMrelic}, this temperature ratio is in turn a proxy for the DM relic abundance. Eliminating the reheaton couplings in favor of the physical quantities $T_{\rm rh}$ and $\xi_{\rm rh}$, one obtains
\begin{equation}\label{eq:reheatonrelation}
    \epsilon_b \gtrsim 5.9\times 10^{-41}\,\frac{\xi_{\rm rh}^2}{\alpha_{\rm rh}^2}\,\frac{M_{\rm P}}{m_\phi}\left(\frac{T_{\rm rh}}{5\,{\rm MeV}}\right)^2\,.
\end{equation}
Moreover, in the case that the dark sector is reheated above the confinement scale, $T_{\rm rh} \xi_{\rm rh} \gtrsim f_\pi$, implies the lower bound
\begin{equation}
    \epsilon_b \gtrsim  1.1\times 10^{-40}\,\frac{M_{\rm P}}{m_\phi}\left(\frac{f_\pi}{5\,{\rm MeV}}\right)^2\,.
\end{equation} 

Summarizing, the explicit $\mathbb{Z}_N$ breaking responsible for the axion potential is not arbitrary. Rather, its magnitude is fixed by the same parameters that govern reheating and, as shown in the next section, determine the DM abundance. In this way, the additional contribution to the axion potential is directly linked to the cosmological origin of DM. In particular, under the DE scale requirement, the lower bound on $\epsilon_b$ above implies an upper bound on $m_\pi f_\pi$ since $m_\phi < M_{\rm P}$.

\section{Dark matter relic density}
\label{sec:DMrelic}

The dark pions are effectively stable since they are the lightest states in the dark sector and all renormalizable interactions, both within the dark sector and between the dark and visible sectors, preserve the pions' parity $\pi\to -\pi$. As a result, dark pions naturally constitute the DM relic abundance. Their axion-dependent mass then directly generates the DE--DM interaction responsible for the finite-density phantom-crossing mechanism, making the dark-pion scenario the minimal realization of this framework.\\

Before discussing the relic abundance of dark pions, let us briefly comment on the possibility that DM is instead primarily composed of dark baryons. In the minimal symmetric thermal history, efficient baryon-antibaryon annihilation into dark pions suppresses the baryonic relic abundance, requiring dark baryon masses of order $m_B\sim 100\,{\rm TeV}$ to reproduce the observed DM density~\cite{Garani:2021zrr}. As shown in~\cref{app:baryonDM}, such a high confinement scale in turn implies an extremely small $\mathbb{Z}_N$--breaking spurion in order to generate the observed DE scale, forcing the dark pion mass to satisfy $m_\pi\lesssim 10^{-5}\,$eV. The resulting suppression of the dark-baryon sigma term effectively decouples the DM and DE sectors, thereby preventing the phantom-crossing mechanism discussed in the previous section. Alternatively, one may consider asymmetric dark baryons. However, as discussed  in~\cref{app:baryonDM}, reproducing the observed DM relic abundance while maintaining an appreciable DE--DM coupling requires a large primordial dark baryon asymmetry, while the associated dark pions remain more abundant and are typically lighter than $\mathcal{O}(10)\,$MeV, necessitating an efficient depletion mechanism compatible with BBN and CMB constraints. For these reasons, in contrast to Ref.~\cite{Khoury:2025txd}, we focus in the remainder of this work on the minimal scenario in which dark pions constitute the dominant DM component.\\

The DM relic abundance of dark pions depends qualitatively on the dark sector phase at the end of reheating. Here, we distinguish two scenarios depending on whether the dark sector is reheated in the deconfined phase, $T'_{\rm rh} > T'_c$, or in the confined phase with $m_\pi\lesssim T'_{\rm rh} < T'_c$.  

In the first case, when the dark sector temperature reaches $T'_c\sim f_\pi$, the confinement phase transition occurs, and the dark degrees of freedom reorganize adiabatically into dark pions and baryons. At the end of the confinement transition, the dark pions are relativistic
provided \(T_c'\gtrsim m_\pi\). Their yield is then
\begin{equation}
    Y_\pi(T_c')
    \equiv \frac{n_\pi(T_c')}{s_{\rm SM}(T_c)}
    \simeq
    0.077\,\xi_c^3
    \left[\frac{10.75}{g_{*s}^{\rm SM}(T_c)}\right],
\end{equation}
where \(\xi_c=T_c'/T_c\). Entropy conservation in the separately reheated visible and dark sectors relates this to the reheating temperature ratio as
\begin{equation}
    \xi_c^3
    =
    \xi_{\rm rh}^3
    \left[
        \frac{g_Dg_{*s}^{\rm SM}(T_c)}
             {g_\pi g_{*s}^{\rm SM}(T_{\rm rh})}
    \right].
\end{equation}

Subsequent number-changing reactions, in particular $4\pi\to2\pi$ in the case of two light flavors\footnote{There are no Wess-Zumino-Witten interactions~\cite{Wess:1971yu} for two light flavors.}, may
further reduce the comoving pion number if they remain efficient after the
pions become nonrelativistic~\cite{Carlson:1992fn}.
We parameterize this possible cannibal
depletion by a factor ${\cal D}_{\rm can}\leqslant 1$, which is explicitly calculated in~\cref{app:cannibal}. We find that the depletion is maximal for $m_\pi\sim f_\pi\sim 10\,$keV, with $D_{\rm can}^{\rm max}\approx 5.5\times 10^{-3}$, while it becomes rapidly suppressed in the chiral regime $m_\pi\ll f_\pi$.

Once chemical decoupling occurs, \(\Gamma_{4\pi\to2\pi}\lesssim H\), the pion number freezes out, and the comoving abundance remains constant until today. The resulting DM relic abundance is therefore~\cite{Gondolo:1990dk}
\begin{equation}
    \Omega_\pi h^2
    \simeq
    0.12
    \left(\frac{m_\pi}{10\,{\rm keV}}\right)
    \left(\frac{\xi_{\rm rh}}{0.036}\right)^3
    \left[
        \frac{10.75}{g_{*s}^{\rm SM}(T_{\rm rh})}
    \right]
    {\cal D}_{\rm can} .
    \label{eq:Ompi}
\end{equation}
In the absence of cannibal depletion, \({\cal D}_{\rm can}=1\), the observed DM relic density is reproduced for  
\(\xi_{\rm rh}=0.036\). Conversely, assuming maximal cannibal depletion
with \({\cal D}_{\rm can}\simeq{\cal D}_{\rm can}^{\rm max}\), the same abundance requires \(\xi_{\rm rh}\simeq0.20\).\\

In the case where reheating occurs below confinement but before the pions become nonrelativistic, $m_\pi \lesssim T'_{\rm rh} \lesssim T_c' \sim f_\pi$, the dark sector is populated directly in the pion effective theory. In this case, there is
no entropy transfer from deconfined dark quarks and gluons. The energy injected by reheaton decays is expected to thermalize efficiently through elastic pion scatterings, leading to a Bose-Einstein distribution
\begin{equation}
    f_\pi(E)=
    \frac{1}{\exp[(E-\mu_\pi)/T']-1},
\end{equation}
with temperature $T'$ and, in general, a nonzero chemical potential \(\mu_\pi\). 
The relativistic pion yield at the end of reheating is therefore
\begin{equation}
    Y_\pi(T'_{\rm rh})
    \simeq
    0.077{\cal C}_{\mu_\pi}\,\xi_{\rm rh}^3
    \left[
        \frac{10.75}{g_{*s}^{\rm SM}(T_{\rm rh})}
    \right]\,,
\end{equation}
where $\mathcal{C}_{\mu_\pi}\equiv{\rm Li}_3(e^{\mu_\pi/T'})/\zeta(3)$, with Li$_3$ the polylogarithm of order 3. Chemical equilibrium corresponds to $\mu_\pi=0$,  and requires efficient number-changing processes, such the $4\pi\leftrightarrow 2\pi$ reactions discussed above, or alternatively a sufficiently strong fragmentation cascade during reheaton decay into the dark sector. If this condition is
not satisfied, the pion number is conserved from reheating onward, and the final abundance depends on the details of the nonthermal production history through $\mu_\pi\neq0$. The corresponding relic abundance is therefore
\begin{equation}
    \Omega_\pi h^2
    \simeq
    0.12
    \left(\frac{m_\pi}{10\,{\rm keV}}\right)
    \left(\frac{\xi_{\rm rh}}{0.083}\right)^3
    \left[
        \frac{10.75}{g_{*s}^{\rm SM}(T_{\rm rh})}
    \right]
    {\cal C}_{\mu_\pi}{\cal D}_{\rm can}\,,
    \label{eq:Ompi-pion-rh}
\end{equation}
where $\cal{D}_{\rm can}$ accounts for possible cannibal depletion when number-changing reactions maintain chemical equilibrium after reheating, so that ${\cal C}_{\mu_\pi}=1$, and remain active into the nonrelativistic regime.\\

Finally, although the reheaton does not decay to the $k\neq0$ sectors, those are nonetheless reheated (in a $\mathbb{Z}_N$--symmetric way) by the SM fields through the freeze-in mechanism~\cite{Hall:2009bx} mediated by gravitational interactions~\cite{Garny:2015sjg}. As a result, the temperature ratio in the $k\neq 0$ sectors is 
$\xi_{\rm rh}^{k\neq 0}\approx 0.34(T_{\rm rh}/M_{\rm P})^{3/4}$~\cite{Redi:2020ffc}.
This gravitational contribution to the relic density is negligible unless the SM is reheated to very high temperatures close to the Planck scale.

\section{Phenomenological constraints}
\label{sec:alltogether}
We now summarize the main phenomenological and consistency constraints on the model. The $\mathbb{Z}_N$--axion framework is characterized by the fundamental parameters $f$, $z_{ud}$, $m_\pi$, $f_\pi$, and $N$, supplemented by a small explicit $\mathbb{Z}_N$--breaking spurion $\epsilon_b$. As discussed in the previous sections, the explicit breaking must reproduce the observed DE scale, $\epsilon_b m_\pi^2f_\pi^2\sim{\rm meV}^4$, while the number of copies $N$ is chosen sufficiently large that the residual $\mathbb{Z}_N$--symmetric axion potential remains subdominant. Moreover, the phantom-crossing scenario selects a restricted region of the parameter space, corresponding to $z_{ud}/f^2\sim \mathcal{O}(10)/M_{\rm P}^2$. 
The remaining low-energy parameters, $m_\pi$ and $f_\pi$, are further constrained by the cosmological thermal history and by DM phenomenology.\\

We begin with constraints arising from late-time DM properties. Structure formation constrains the dark-pion velocity dispersion rather than the mass alone. Matching the velocity dispersion of our three real bosonic pion relic to that of a standard thermal WDM relic gives
$m_\pi\simeq0.75\,m_{\rm WDM}^{\rm th}$~\cite{Bode:2000gq}. Current Lyman-$\alpha$ constraints on thermal WDM~\cite{Viel:2013fqw} therefore translate into a lower bound of the order of a few keV on the dark-pion mass. In the following, we adopt  the conservative requirement
\begin{equation}
m_\pi\gtrsim10\,{\rm keV}.
\end{equation}
Combined with the DM relic density constraint, this lower limit further implies $\xi_{\rm rh}\lesssim0.1$, consistent with the expectation that the dark sector is reheated less efficiently than the visible sector, corresponding to $\kappa_D<\kappa_{\rm SM}$.

Dark-pion self-interactions provide a complementary constraint. Observations of colliding galaxy clusters require approximately $\sigma_{\rm self}/m_{\rm DM}\lesssim 1\,{\rm cm}^2/{\rm g}$~\cite{Markevitch:2003at}, while the nonrelativistic dark-pion scattering cross-section scales parametrically as $\sigma_{\pi\pi}\simeq c_{\pi\pi} m_\pi^2/(64\pi f_\pi^4)$, with $c_{\pi\pi}=23/6\simeq3.8$ at leading order in chiral perturbation theory~\cite{Weinberg:1966kf}. This translates into the lower bound,
\begin{equation}
    f_\pi\gtrsim 2.5\,{\rm MeV} \left(\frac{c_{\pi\pi}}{23/6}\right)^{1/4}\left(\frac{10\,{\rm keV}}{m_\pi}\right)^{1/4}\,,
\end{equation}
which implies $m_\pi\ll 4\pi f_\pi$, ensuring the validity of the chiral effective theory.\\

We now turn to the constraints imposed by the cosmological thermal history. First, the dark sector must be reheated above the dark-pion mass, $T_{\rm rh}'=T_{\rm rh}\xi_{\rm rh}>m_\pi$. As discussed in~\cref{sec:ZNbreaking}, the reheaton couplings determine both the visible- and dark-sector reheating temperature, as well as the size of the $\mathbb{Z}_N$--breaking spurion. Combining~\cref{eq:Trh,eq:Thratio,eq:reheatonrelation} and imposing the observed DE scale yields 
\begin{equation}\label{eq:Trhprime}
 T_{\rm rh}'\simeq 46\,{\rm GeV}\,\sqrt{\frac{m_\phi}{M_{\rm P}}}\left(\frac{10\,{\rm keV}}{m_\pi}\right)\left(\frac{2\,{\rm MeV}}{f_\pi}\right)\,.   
\end{equation}
Requiring $T_{\rm rh}'>m_\pi$ then implies a upper bound on $f_\pi$,
\begin{equation}
    f_\pi\lesssim 9.1\,{\rm TeV}\sqrt{\frac{m_\phi}{M_{\rm P}}}\left(\frac{10\,{\rm keV}}{m_\pi}\right)^2\,.
\end{equation}

Successful BBN further requires sufficiently efficient reheating of the visible sector, namely $T_{\rm rh}\gtrsim 5-10\,{\rm MeV}$. Combining~\cref{eq:Trhprime} with the DM relic density constraint in~\cref{eq:Ompi-pion-rh} yields
\begin{equation}
    T_{\rm rh}\simeq 260\,{\rm GeV}\sqrt{\frac{m_\phi}{M_{\rm P}}}\left(\frac{10\,{\rm keV}}{m_\pi}\right)^{2/3}\left(\frac{2\,{\rm MeV}}{f_\pi}\right)\,,
\end{equation}
where we have assumed $\mathcal{C}_{\mu_\pi}=\mathcal{D}_{\rm can}=1$, $\Omega_\pi h^2=0.12$ and $g_\star^{\rm SM}(T_{\rm rh})=106.75$. Imposing the BBN constraint then gives
\begin{equation}
f_\pi\lesssim 100\,{\rm GeV}\sqrt{\frac{m_\phi}{M_{\rm P}}}\left(\frac{10\,{\rm keV}}{m_\pi}\right)^{2/3}\left[\frac{10.75}{g_\star^{\rm SM}(T_{\rm rh})}\right]^{1/3}
\end{equation}
For values of $f_\pi$ saturating this bound one finds $\mathcal{D}_{\rm can}\simeq 1$, justifying a posteriori the approximation adpoted above.\\

\begin{figure}[t]
    \centering
    \includegraphics[width=\columnwidth]{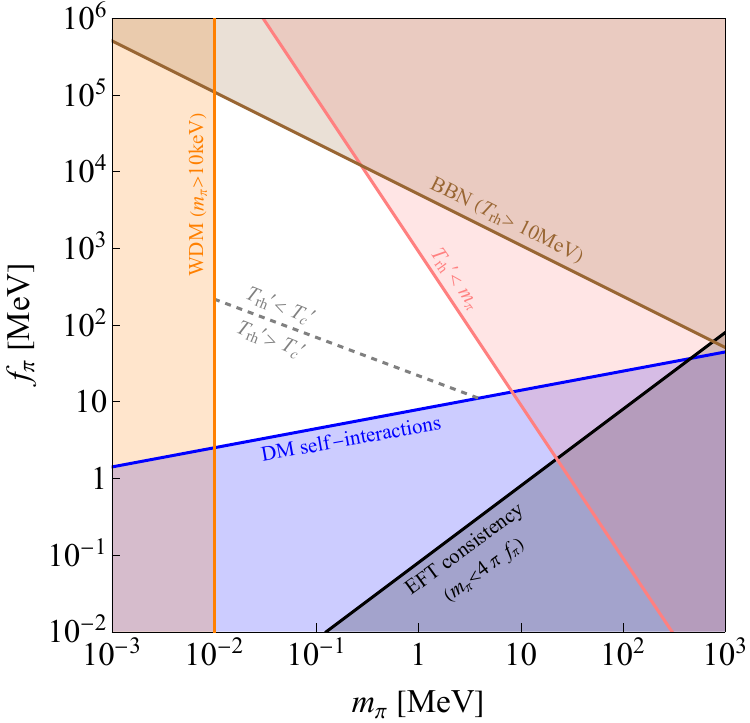}
    \caption{
    Phenomenological constraints from cosmology and astrophysics in the \((m_\pi,f_\pi)\) plane, including bounds from WDM (orange), DM self-interactions (blue), BBN (brown), the consistency of the pion EFT (black) and  after imposing the requirement of reheating the dark sector above $m_\pi$ (pink). The corresponding shaded regions are excluded, under the conditions that the observed DE scale and DM relic abundance are reproduced, and that the reheaton mass $m_\phi\leqslant M_P$. Above (below) the dashed gray line, the dark sector is reheated in the confined phase with relativistic dark-pions (deconfined phase). See~\cref{sec:alltogether} for details. 
    }
    \label{fig:parameter-space}
\end{figure}

The resulting viable parameter space in the $(m_\pi,f_\pi)$ plane is shown in~\cref{fig:parameter-space}, assuming $m_\phi=M_{\rm P}$. The lower boundaries are set by bounds from WDM free-streaming and dark-pion self-interactions, while the upper boundaries follow from the requirement of a consistent reheating history exists. We emphasize that decreasing the reheaton mass progressively shrinks the allowed region, which disappears entirely for $m_\phi\lesssim 10^9\,$GeV.

In summary,~\cref{fig:parameter-space} demonstrates that a sizeable region of parameter space simultaneously accommodates the observed DM abundance and  DE scale, while remaining compatible with a consistent cosmological history originating from reheaton decays.

\section{Conclusions}
\label{sec:concl}

A confirmed preference for an effective dark-energy equation of state crossing the phantom divide would present a major challenge for canonical quintessence models. In this work, building on the mechanism proposed in Ref.~\cite{Khoury:2025txd}, we have presented a microscopic realization of apparent phantom dark energy based on an axion coupled to a confining dark sector. The key observation is that a finite density of DM can initially trap the axion near a metastable configuration, while the subsequent dilution of the DM density releases the field and induces a late-time evolution that mimics an effective equation of state \(w_{\rm DE}^{\rm eff}<-1\), without introducing ghosts or violating the null-energy condition.\\

The model relies on an axion coupled to \(N\) copies of a two-flavor
dark QCD sector related by a discrete \(Z_N\) exchange symmetry. This symmetry exponentially suppresses the axion potential in a technically natural manner. However, it also reduces the periodicity of the potential, thereby preventing the large-field excursion required for a realistic cosmological evolution. 
Remarkably, we have shown that this obstruction can be overcome through a controlled explicit breaking of \(Z_N\),  induced by reheating into a single dark sector.
Reheating then simultaneously populates the Universe with dark-pion DM from that selected copy and restores the physical \(2\pi f\)-periodic axion potential. The finite density of these dark pions initially traps the axion near a
density-induced minimum, while the vacuum potential generated by the reheating-induced
$\mathbb{Z}_N$--breaking selects a different minimum. As the Universe expands and the dark-pion
density redshifts away, the axion is released and rolls toward the true vacuum (\cref{fig:axionsolution}). During this evolution, the dark-pion mass increases, generating the energy
transfer responsible for the apparent phantom behavior (\cref{fig:wDEplot}). 

A central feature of the construction is that the ingredients responsible for the late-time DE dynamics are directly tied to the early thermal history of the Universe. The reheaton coupling that selects the populated dark sector also generates the small \(Z_N\)-breaking spurion controlling the vacuum axion potential. As a result, the DE scale, the dark-pion relic abundance, and the reheating history are not independent ingredients, but are linked by a predictive set of parametric relations involving
\(m_\pi\), \(f_\pi\), and the reheating temperatures of the dark and visible sectors. We have identified a finite region of parameter space in which these relations are simultaneously compatible with standard BBN, WDM free-streaming constraints, DM self-interaction bounds, and the validity of the chiral effective theory (\cref{fig:parameter-space}). In particular, we find that the regime in which the selected dark sector is reheated above confinement provides a minimal, internally consistent, and technically natural realization of the finite-density phantom mechanism with dark-pion DM.

While the benchmark solution presented in this work reproduces the qualitative pattern suggested by current DESI reconstructions, namely a transition from $w_{\rm DE}^{\rm eff}\gtrsim-1$ today to $w_{\rm DE}^{\rm eff} <-1$ in the recent past, a dedicated likelihood analysis including BAO, SNe, CMB, weak-lensing, and growth measurements (\cref{fig:growth}) is required to assess quantitatively the preference of the model over $\Lambda$CDM. Such an analysis would be particularly interesting because the present framework predicts a correlated pattern of deviations in the expansion history, the effective DE equation of state, and the growth of structure, rather than an arbitrary phenomenological parametrization of DE. Constructing model-specific fitting templates, for example in terms of the release redshift, the depth of the transient phantom phase, and the strength of the DM--DE interaction, would therefore provide a natural interface between the microscopic theory and cosmological observations. On the theoretical side, it would be interesting to explore the dark thermal history, including the role of cannibalization across the viable parameter space, and to investigate explicit UV completions in which the origin of the large discrete symmetry and the axion quality are manifest. More generally, the framework developed here provides a concrete example of how an apparent phantom equation of state may emerge from technically natural and microscopically consistent quantum field theory dynamics.

\section*{Acknowledgments}

We acknowledge the Moriond Cosmology 2026 conference, particularly Justin Khoury's talk, which sparked this work. We thank ChatGPT for many fruitful technical discussions and Wolfram Mathematica for patiently solving the coupled equations after we stopped trying to do so analytically.

\appendix

\section{DE diagnostics}\label{app:DEdiagnostics}

In this appendix, we collect additional background-level diagnostics of the
DE evolution for the benchmark solution discussed in the main text.
These quantities provide complementary diagnostics of the background expansion beyond the effective equation of state shown in
\cref{fig:wDEplot}. 

We first consider the effective DE density normalized to its present value, 
\begin{equation}
    f_{\rm DE}(z)\equiv \frac{\rho_{\rm DE}(z)}{\rho_{\rm DE}(0)}\,,
\end{equation}
which directly relates to the equation of state $w_{\rm DE}^{\rm eff}$,
\begin{equation}
    \frac{1}{3}\log f_{\rm DE}(z)=\int_0^z \frac{dz'}{1+z'}\left[1+w_{\rm DE}^{\rm eff}(z')\right]\,.
\end{equation}
Other useful variables are the $Om$ diagnostic,
\begin{equation}
    Om(z)\equiv \frac{h^2(z)-1}{(1+z)^3-1}\,,
\end{equation}
and the deceleration parameter,
\begin{equation}
     q(z)\equiv \frac{d\log h}{d\log (1+z)}-1\,,
\end{equation}
which offer robust tests of departures from $\Lambda$CDM. Both depend only on the shape of the expansion history through the normalized Hubble parameter $h(z)\equiv H(z)/H_0$, making them largely insensitive to degeneracies in the absolute dark-energy and matter densities. In particular, ${\rm Om}(z)$ is constant in $\Lambda$CDM, whereas $q(z)$ directly tracks the transition from decelerated to accelerated expansion.\\
\begin{figure}
    \centering
    \includegraphics[width=\linewidth]{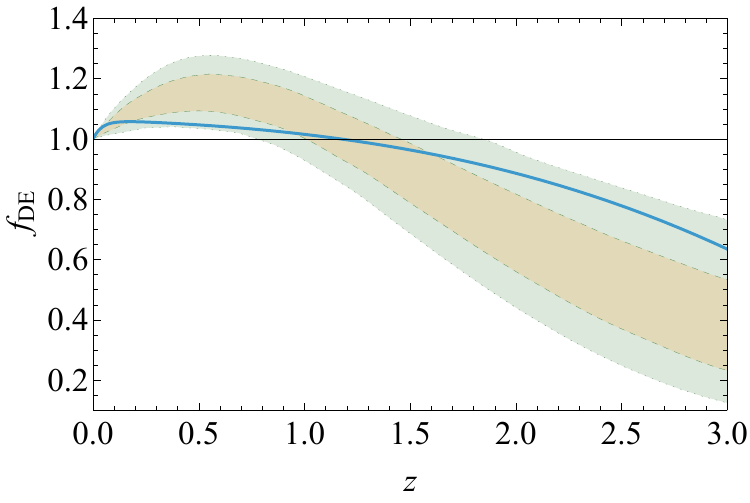}\\
    \includegraphics[width=\linewidth]{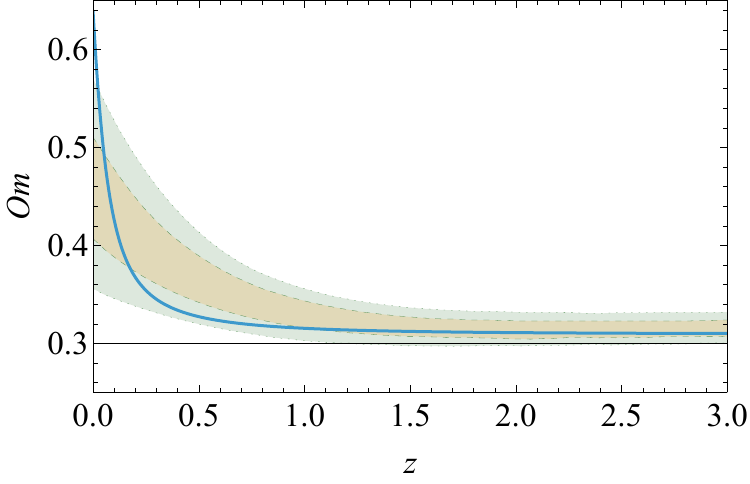}\\
    \includegraphics[width=\linewidth]{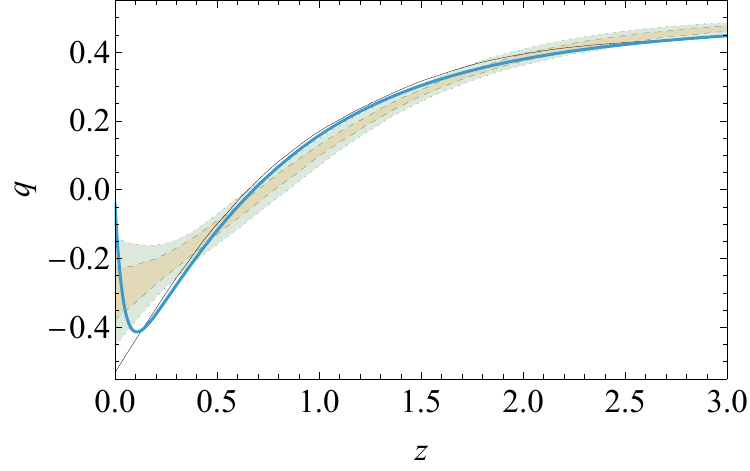}
    \caption{Additional background diagnostics for the benchmark solution shown in the main text (blue line). Top: normalized dark-energy density $f_{\rm DE}$. Middle: $Om$ diagnostic, equal to $\Omega_m=0.302$ in the $\Lambda$CDM model. Bottom: deceleration parameter $q$. The shaded dark and light bands indicate the $68\%$ and $95\%$ confidence regions, respectively, obtained from Gaussian process reconstructions of $w_{\rm DE}(z)$ using the combined DESI, CMB, and Union3 datasets~\cite{DESI:2025fii}. Together these observables highlight the correlated departures from the $\Lambda$CDM expansion history induced by the late-time release of the axion.}
    \label{fig:de-diagnostics}
\end{figure}

\Cref{fig:de-diagnostics} shows the evolution of these diagnostics for the benchmark solution presented in the main text. Consistent with the DESI reconstruction~\cite{DESI:2025fii}, the effective DE density exceeds its present value at intermediate redshifts, with a maximum marking crossing of the effective phantom divid. This behavior reflects the gradual transfer of energy from DE to the DM sector during the late-time axion evolution. The same dynamics produce correlated deviations of both $Om(z)$ and $q(z)$ from their $\Lambda$CDM predictions over the redshift range in which the transient phantom phase occurs. Also, the benchmark predicts a rapid recent weakening of cosmic acceleration, with $q(z)$ turning upward and approaching zero at the present epoch. At larger redshifts, the effective DE contribution becomes dynamically negligible, and both $Om(z)$ and $q(z)$ approach their matter-dominated behavior. These diagnostics therefore provide a complementary visualization of the same microscopic mechanism responsible for the evolving DE equation of state.


\section{Baryonic DM}\label{app:baryonDM}

Here, we focus on the case where the dark sector is reheated in the deconfined phase, $T'_{\rm rh}>T_c'\sim f_\pi$. At the confinement transition, the dark quarks reorganize into dark pions and baryons. Although the lightest dark baryons are stable due to dark baryon number conservation, their abundance is already strongly suppressed at confinement because their mass, $m_B\sim\Lambda_D\sim 4\pi f_\pi$, exceeds the phase-transition temperature. As a result, the initial baryon-to-pion ratio is Boltzmann-suppressed, 
\begin{equation}
\frac{n_B}{n_\pi}\simeq 0.7\left(\frac{m_B}{T'_c}\right)^{3/2}e^{-\frac{m_B}{T'_c}}\sim\mathcal{O}(10^{-4})\,,  
\end{equation} 
corresponding to a dark-baryon (and antibaryon) yield
\begin{align}
    Y_{B}(T'_c)\simeq 8.4\times 10^{-7}\xi_c^3\left[\frac{106.75}{g^{\rm SM}_{\star s}(T_c)}\right]\,.
\end{align}
In the absence of a primordial dark-baryon asymmetry, subsequent baryon-antibaryon annihilations into dark pions further suppress the baryonic abundance. As a result, reproducing the observed DM abundance requires a large dark-baryon mass~\cite{Garani:2021zrr}, 
\begin{equation}
    \Omega_Bh^2\simeq 0.12\left[\frac{g_{\star s}^{\rm SM}(T_f)}{106.75}\right]^{1/2}\left(\frac{m_B}{100\,{\rm TeV}}\right)^2\left(\frac{\xi_{\rm rh}}{0.42}\right)\,,
\end{equation}
where $T_f\sim m_B/25$ denotes the SM temperature at baryon freeze-out. 

Such a heavy baryon implies an extremely small $\mathbb{Z}_N$--breaking spurion in order to reproduce the DE scale, $\epsilon_bm_\pi^2f_\pi^2\sim {\rm meV}^4$, namely
\begin{equation}
 \epsilon_b\simeq 1.9\times10^{-46}\,\xi_{\rm rh}\left(\frac{10\,{\rm keV}}{m_\pi}\right)^2\,,
\end{equation}
where we assumed $\Omega_Bh^2=0.12$ and $g_{\star s}^{\rm SM}(T_f)=106.75$. However, as discussed in~\cref{sec:ZNbreaking}, the reheating framework imposes a lower bound on $\epsilon_b$, 
\begin{equation}
\epsilon_b\gtrsim 2.4\times10^{-28}\frac{M_{\rm P}/m_{\phi}}{\xi_{\rm rh}}\left(\frac{T_{\rm rh}}{T_{\rm rh}^{\rm min}}\right)^2\,,
\end{equation}
where $T_{\rm rh}^{\rm min}\sim m_B/(4\pi\xi_{\rm rh})$ is the minimal SM temperature required to reheat the dark sector above the confinement scale. Combining these constraints yields
\begin{equation}
    m_\pi\lesssim 8.9\times 10^{-6}\,{\rm eV}\,\frac{\xi_{\rm rh}}{\sqrt{M_{\rm P}/m_{\phi}}}.
\end{equation}
Such an ultralight dark pion implies an extremely suppressed coupling between dark baryons and the axion field. Indeed, this coupling is controlled by the dark-baryon sigma term, $\sigma_B=m_q\partial m_B/\partial m_q$, which vanishes in the chiral limit. Using the above bounds and the GOR relation, one finds $\sigma_B/m_B\sim m_\pi^2/m_B^2\lesssim \mathcal{O}(10^{-38})$. The resulting DM-DE interaction is therefore completely negligible, preventing the phantom-crossing mechanism discussed in the main text.\\

Asymmetric dark baryons  provide a possible alternative realization of the DM sector. Assuming a primordial asymmetry $\eta \equiv 1-n_{\bar{B}}(T'_c)/n_B(T'_c)>0$ generated prior to dark confinement, the surviving asymetric relic abundance after baryon-antibaryon annihilation is 
\begin{equation}\label{eq:OmBasym}
 \Omega_B^ah^2\simeq 1.2\times 10^4\eta \frac{\xi_{\rm rh}^3}{g_{\star s}^{\rm SM}(T_c)}\left(\frac{m_B}{40\,{\rm MeV}}\right)\,,
\end{equation}
where $m_B\ll 100\,{\rm TeV}$ so that the symmetric component annihilates efficiently.
Imposing the DE scale constraint and the lower bound on $\epsilon_b$ in~\cref{eq:reheatonrelation}, the DM relic abundance $\Omega_B^ah^2=0.12$ requires
\begin{equation}\label{eq:Basym}
\eta\gtrsim \frac{9.8\times 10^{-5}}{\xi_{\rm rh}^3}\left(\frac{\sigma_B}{m_B}\right)^{1/6}\left(\frac{T_{\rm rh}}{T_{\rm rh}^{\rm min}}\right)^{1/3}\left(\frac{M_{\rm P}}{m_\phi}\right)^{1/6}\,,
\end{equation}
assuming $g_{\star s}^{\rm SM}(T_c)=10.75$. Since $\xi_{\rm rh}\lesssim1$ and $m_\phi\lesssim M_{\rm P}$, and restricting to $\sigma_B/m_B\gtrsim 10^{-3}$ to generate an appreciable DM-DE interaction~\cite{Khoury:2025txd}, the required dark-baryon asymmetry $\eta$ is many orders of magnitude larger than the observed SM baryon asymmetry, $\eta_{\rm SM}\sim\mathcal{O}(10^{-9})$. Therefore, the asymmetric baryonic DM scenario necessitates a dedicated mechanism for generating the large dark asymmetry in~\cref{eq:Basym}, most likely of the Affleck-Dine type~\cite{Affleck:1984fy}. 
If dark pions were effectively stable, they would contribute to the DM relic density, with abundance
\begin{equation}
    \Omega_\pi h^2\simeq\frac{9.4\times 10^2}{\eta}\mathcal{D}_{\rm can}\left(\frac{\sigma_B/m_B}{10^{-3}}\right)^{1/2}\,.
\end{equation}
Even assuming $\eta\sim\mathcal{O}(1)$, 
dark pions would still dominate the DM. Hence,  additional interactions must be introduced to decay the dark pions back to the SM. Combining~\cref{eq:OmBasym} and~\cref{eq:Basym} implies an upper bound on the dark baryon mass,
\begin{equation}
    m_B\lesssim 130\,{\rm MeV}\left(\frac{10^{-3}}{\sigma_B/m_B}\right)^{1/6}\left(\frac{T_{\rm rh}^{\rm min}}{T_{\rm rh}}\right)^{1/3}\left(\frac{m_\phi}{M_{\rm P}}\right)^{1/6}\,,
\end{equation}
and, since typically $m_\pi\lesssim f_\pi\sim m_B/(4\pi)$, the dark-pion mass must satisfy $m_\pi\lesssim 10\,{\rm MeV}$.\\

The baryonic DM scenario therefore requires several additional ingredients beyond the minimal framework considered in this work. First, a mechanism capable of generating a large primordial dark-baryon asymmetry is needed. Second, the dark-baryon spectrum must remain sufficiently light to ensure an appreciable dark-baryon sigma term and hence a sizeable DM-DE interaction. Third, the accompanying population of light dark pions must be efficiently depleted without violating BBN or CMB constraints.

By contrast, the dark-pion scenario studied in the main text identifies the lightest dark hadrons themselves as the DM relic abundance and directly exploits their axion-dependent mass to generate the DM-DE interaction. It therefore provides the minimal realization of the finite-density phantom-crossing mechanism within the present framework.

\section{Cannibalism }\label{app:cannibal}

When dark pions become non-relativistic, efficient annihilations may deplete their number density. Since the dark sector is secluded from the SM, the only number-changing processes are $n\pi\to m\pi$ with $m<n$ induced by pion self-interactions. If $2\pi\to 2\pi$ scattering is fast enough, the energy of the disappearing $n-m$ pions is efficiently redistributed among the remaining ones, increasing their temperature. This phase is known as cannibalism, where secluded non-relativistic dark-sector particles `eat' themselves to keep warm~\cite{Carlson:1992fn}. For $N_f=2$, the CP symmetry of the dark sector implies that the dominant process is $4\pi\to 2\pi$, whose rate is given by~\cite{Hochberg:2014dra} 
\begin{equation}
    \Gamma_{4\to 2}\equiv n_\pi^3\langle \sigma_{4\to 2}v^3\rangle\,,
\end{equation}
where\footnote{The pion chemical potential $\mu_\pi$ vanishes during the cannibal phase.} 
\begin{equation}
n_\pi(T')\simeq g_\pi\left(\frac{m_\pi T'}{2\pi}\right)^{3/2}\exp\left(-\frac{m_\pi}{T'}\right)\,,
\end{equation}
is the pion number density in the non-relativistic limit ($T'\lesssim m_\pi$), and a parametric estimate of the thermal-averaged cross-section is 
\begin{equation}
    \langle \sigma_{4\to 2}v^3\rangle \equiv \frac{c_{6}}{f_\pi^8}\,,
\end{equation}
where $c_{6}$ is a dimensionless constant. \\ 

The evolution of the dark-sector temperature during the cannibal phase is obtained using the separate conservation of the comoving entropy of the dark sector, yielding approximately~\cite{Carlson:1992fn} 
\begin{equation}
    x'\equiv \frac{m_\pi}{T'}\simeq 3\log\left(\frac{m_\pi}{\xi_c T}\right)\,,
\end{equation}
for $T'<m_\pi$, where we neglected the effect of the decoupling of SM species between $T_c$ and the SM temperature when pions become non-relativistic. Hence, the dark-sector `clock' runs exponentially slower than the SM one during the cannibal phase.\\

The cannibal phase ends when $4\pi\to 2\pi$ processes decouple at $x'=x'_f$ given by $\Gamma_{4\to 2}=H(T)$, yielding
\begin{align}
x'_f\simeq&\ 20.9+\frac{24}{7}\log\left(\frac{ m_\pi}{f_\pi}\right)\nonumber\\
&\ +\frac{3}{7}\log\left[c_6\xi_c^2\sqrt{\frac{10.75}{g_\star^{\rm SM}(T_f)}}\left(\frac{10\,{\rm keV}}{m_\pi}\right)\right]\,,
\end{align}
where $T_f\simeq (m_\pi/\xi_c) e^{-x'_f/3}$ is the SM temperature when $T'=T'_f$, and the final dark pion yield is
\begin{equation}
    Y_\pi(T_f')\equiv\mathcal{D}_{\rm can}Y_\pi(T'_c)\,,
\end{equation}
where the cannibal depletion factor is approximately
\begin{equation}
    \mathcal{D}\simeq 5.5\times 10^{-3}\left[\frac{g^{\rm SM}_\star(T_c)}{g^{\rm SM}_\star(T_f)}\right]\left(\frac{20.9}{x_f'}\right)^{3/2}\,.
\end{equation}

\bibliographystyle{JHEP}
\bibliography{refs.bib}

\end{document}